\documentclass[preprint]{aastex}

\shorttitle{A {\it WISE} view of near-UV and mid-IR-excess galaxies}

\shortauthors{Jongwan Ko et al.}

\begin{document}

\title{THE MID-INFRARED AND NEAR-ULTRAVIOLET EXCESS EMISSIONS OF QUIESCENT GALAXIES ON THE RED SEQUENCE}

\author{Jongwan Ko\altaffilmark{1}, Ho Seong Hwang\altaffilmark{2}, Jong Chul Lee\altaffilmark{1}, 
and Young-Jong Sohn\altaffilmark{3}}

\altaffiltext{1}{Korea Astronomy and Space Science Institute, Daejeon 305-348, Republic of Korea}
\altaffiltext{2}{Smithsonian Astrophysical Observatory, 60 Garden Street, Cambridge, MA 02138, USA}
\altaffiltext{3}{Department of Astronomy, Yonsei University, Seoul 120-749, Republic of Korea}
\email{jwko@kasi.re.kr}

\begin{abstract}

We study the mid-infrared (IR) and near-ultraviolet (UV) excess emissions of 
 spectroscopically selected quiescent galaxies on the optical red sequence.
We use the Wide-field Infrared Survey Explorer ({\it WISE}) mid-IR and 
 Galaxy Evolution Explorer ({\it GALEX}) near-UV data for a spectroscopic sample 
 of galaxies in the Sloan Digital Sky Survey Data Release 7 to study the possible 
 connection between quiescent red-sequence galaxies with and without mid-IR/near-UV excess.
Among 648 12 $\mu$m detected quiescent red-sequence galaxies without H$\alpha$ emission, 
 26\% and 55\% show near-UV and mid-IR excess emissions, respectively.
When we consider only bright ($M_{r}$ $<$ $-$21.5) galaxies with early-type morphology, 
 the fraction of galaxies with recent star formation is still 39\%. 
The quiescent red-sequence galaxies with mid-IR and near-UV excess emissions are optically fainter 
 and have slightly smaller $D_{n}$4000 than those without mid-IR and near-UV excess emissions.
We also find that mid-IR weighted mean stellar ages of quiescent red-sequence galaxies with
 mid-IR excess are larger than those with near-UV excess, and smaller than 
 those without mid-IR and near-UV excess.
The environmental dependence of the fraction of quiescent red-sequence galaxies with 
 mid-IR and near-UV excess seems strong even though the trends of quiescent red-sequence galaxies 
 with near-UV excess differ from those with mid-IR excess. 
These results indicate that the recent star formation traced by near-UV ($\lesssim 1$ Gyr)
 and mid-IR ($\lesssim 2$ Gyr) excess is not negligible among nearby, quiescent, red, early-type
 galaxies.
We suggest a possible evolutionary scenario of quiescent red-sequence galaxies 
 from quiescent red-sequence galaxies with near-UV excess to those with mid-IR excess to those
 without near-UV and mid-IR excess.

\end{abstract}


\section{INTRODUCTION}

It is well known that the distribution of nearby galaxies in the optical color-magnitude diagram 
 is bimodal; quiescent, early-type galaxies populate
 a narrow red sequence and star-forming, late-type galaxies form a big blue 
 cloud (e.g., Strateva et al. 2001; Blanton et al. 2003; Baldry et al. 2004; 
 Balogh et al. 2004; Choi et al. 2007 and references therein). 
%Moreover, the tight red sequence is often used as a tool for studying the formation 
%and evolution of ETGs. For example, ETGs in nearby galaxy clusters inhabit a tight 
%locus with very small scatter in the optical CMDs, suggesting the stellar population 
%of ETGs is thought to be homogeneously old (Bower et al. 1992). 
%This result is often interpreted as evidence for the monolithic scenario (e.g., Larson 1974).

Recent studies however reveal that the optical red sequence contains not only early-type galaxies
 with signs of recent star formation (e.g., Yi et al. 2005; Bressan et al. 2006; 
 Schawinski et al. 2007; Clemens et al. 2009; Ko et al. 2009; Lee et al. 2010; Vega et al. 2010; 
 Ko et al. 2012), but also late-type galaxies with optical colors reddened by dust extinction or 
 with low level of star formation (e.g., Bamford et al. 2009; Gallazzi et al. 2009; 
 Wolf et al. 2009; Masters et al. 2010; Ko et al. 2012).
The presence of various types of galaxies on the red sequence is easily found when one
 examines these red-sequence galaxies at different wavelengths. The near-ultraviolet (UV) 
 observations show a diversity of early-type galaxies depending on the amount of recent star 
 formation (e.g., Ferreras \& Silk 2000; Yi et al. 2005; Schawinski et al. 2007; Kaviraj et al.  
 2007). 
The mid-infrared (IR) observations also show that a significant fraction of nearby early-type  
 galaxies has excess flux over the photospheric emission (e.g., Bressan et al. 2006; Clemens et 
 al. 2009; Ko et al. 2009, 2012; Shim et al. 2011; Hwang et al. 2012).

The optical colors of galaxies can be red enough to join the red sequence just 1$-$2 Gyr 
 after the stop of major episodes of star formation (e.g., Bower et al. 1998; Poggianti et al.  
 1999). 
However, previous studies of early-type galaxies on the red sequence showed the large scatter in 
 the near-UV--optical color-magnitude diagrams, which cannot be explained without some amounts of 
 recent star formation (e.g., Ferreras \& Silk 2000; Deharveng et al. 2002). Thanks to the Galaxy 
 Evolution Explorer ({\it GALEX}; Martin et al. 2005), it is generally accepted that the near-UV--
 optical color can be an excellent indicator of recent star formation ($\lesssim$ 1 Gyr) even for 
 a small amount of star formation (e.g., $\sim$1\% of stellar mass). 
This is mainly because the near-UV light is much more sensitive to young stars than the
 bluest optical band. 
Balmer absorption lines are also useful indicators of recent star formation because 
 they trace intermediate-age stars (strongest for A-type stars) that dominate the light 
 of a galaxy 1$-$1.5 Gyr after active star formation stops (Dressler \& Gunn 1983; 
 Couch \& Sharples 1987).
Recently, Choi et al. (2009) studied the possible connection of the UV-excess early-type galaxies 
 to E+A galaxies (i.e., post-starburst) using $\sim$1000 E+As selected from the Sloan Digital Sky 
 Survey (SDSS; York et al. 2000) data. %and $\sim$20 000 SDSS ETGs with {\it GALEX} near-UV data. 
They found that the post-starburst systems such as E+As are characterized by the UV excess 
 without H$\alpha$ emission. 
They suggested that the Balmer absorption line used for identifying E+As can be replaced by 
 near-UV--optical colors that are much sensitive to recent star formation.
%and argued that those represents old E+As (i.e. the final phase of the E+A to the red sequence). 

The mid-IR emission from dusty circumstellar envelopes of asymptotic giant branch (AGB) stars is 
 an also useful indicator of star formation history because it traces stars with ages from a few  
 0.1 Gyr up to the Hubble time; low- to intermediate-mass stars evolve into the AGB phase and 
 their mid-IR emissions decrease with stellar age.
Frogel et al. (1990) found that the contribution of AGB stars to the bolometric luminosity could
 be larger than 40\% at ages from 1.1 to 3.3 Gyr, but rapidly decreases down to 5\% at 10 Gyr. 
Recent stellar population models (e.g., Maraston 2005; Bruzual 2009) suggest that AGB stars with 
 0.1$-$1.5 Gyrs can contribute up to $\sim$ 50\% of the $K$-band light in galaxies.
%From recent population synthesis models (e.g., Maraston 2005; Bruzual 2009), the $K$-band light 
%is produced by AGB stars (mainly carbon stars) up to $\sim$ 50\% at 0.1$-$1.5 Gyr. 
Although the understanding of AGB dust envelopes and their contribution to the mid-IR emission 
 in a galaxy is far from complete (e.g., Marigo et al. 2008), recent mid-IR observations show 
 that a significant fraction of early-type galaxies on the red-sequence have excess flux over  
 photospheric emission (Ko et al. 2009, 2012; Shim et al. 2011; Hwang et al. 2012). 
Unusual polycyclic aromatic hydrocarbons (PAHs) are also detected in the mid-IR spectra of 
 early-type galaxies (e.g., Kaneda et al. 2008; Vega et al. 2010; Panuzzo et al. 2011)
 even though the PAH emission is a typical feature of star-forming galaxies (e.g., Peeters et al. 
 2004). 
This unusual PAHs in early-type galaxies seems to result from because of
 material continuously released by intermediate-age carbon stars; these stars formed in a 
 burst of star formation that occurred within the last few Gyrs (e.g., Vega et al. 2010). 
In summary, if the circumstellar envelopes of AGB stars are the main source of mid-IR emission 
 in galaxies, the mid-IR emission is sensitive to star formation over relatively long ($\gtrsim$ 1  
 Gyr) timescales compared to spectroscopic star formation indicators such as nebular emission 
 lines or Balmer absorption lines (Piovan et al. 2003; Salim et al. 2009; Kelson \& Holden 2010; 
 Donoso et al. 2012).

%From a photometric point, the near-UV and mid-IR light both can trace recent SF history of the galaxy, but
%thier physical source is different and their correlation have subtle complications.
Around 1 Gyr after a single-burst of star formation, massive stars (O, B, and A stars) will be 
 expired. The near-UV flux or H$\beta$ line index then is no longer a good tracer of star 
 formation history in galaxies. 
However, the mid-IR emission can trace star formation over much longer timescales because low to 
 intermediate mass (1 $-$ 9 M$_{\odot}$) stars evolve to the AGB phase, and their 
 circumstellar dust emission is strong in the mid-IR. Therefore, the near-UV continuum 
 and the mid-IR emission are complementary each other, and can provide a complete view of 
 recent star formation history of galaxies.

In this paper, we study the various types of galaxies on the red sequence 
(i.e., optically quiescent without ongoing star formation) using near-UV and mid-IR data to 
understand the possible connection between them. Section 2 describes the observational data 
we use. We examine the near-UV and mid-IR properties of quiescent red-sequence galaxies 
in Section 3, and conclude in Section 4. 
Throughout, we use the AB magnitude system, and adopt flat $\Lambda$CDM 
cosmological parameters: $H_0 = 70$ km s$^{-1}$ Mpc$^{-1}$, $\Omega_{\Lambda}=0.7$ and 
$\Omega_{m}=0.3$.

%ETGs with excess mid-infrared (mid-IR) emission from dusty circumstellar envelopes of asymptotic
%giant branch (AGB) stars, though the emission can also originate from active galactic nuclei (AGNs).
%(e.g., Faber et al. 2007; Lee et al. 2008; Ko et al. 2009; 
%Cortese \& Hughes 2009; Wolf et al. 2009; Gallazzi et al. 2009; Tran et al. 2009; 
%Bamford et al. 2009; Bundy et al. 2010; Masters et al. 2010; Salim \& Rich 2010; Ko et al. 2012). 

%Recent {\it GALEX} studies of nearby ETGs interpret this large scatter in the near-UV CMR as recent SFA,
%since the near-UV light is more sensitive to the young stars than the optical light and less sensitive 
%to UV upturn phenomenon than the far-UV (e.g., Ferreras et al. 2002; Yi et al. 2005 REF more!!).

%We expect that there are two different phases (star formation quenching and morphology 
%changing), when a blue, SF LTG turns into a red, quiescent ETG. Recent studies revealed 
%that SF quenching (optical color change) is not always accompanied by morphological change 
%(e.g., Blanton et al. 2005, S\'{a}nchez et al. 2007, Bamford et al. 2009; Wolf et al. 2009). 
%In other words, the time scale of transformation from blue to red and of morphological 
%change from late-type to early-type is different, and seems to be a function of stellar 
%mass and environment. Observationally, the existence of red or passive spiral galaxies  
%and blue ETGs, and their preference for specific masses and local densities 
%supports this idea (e.g., Bamford et al. 2009; Wolf et al. 2009; Masters et al. 2010; Ko et al. 2012) 

\section{THE DATA AND THE SAMPLE}

\subsection{Galaxy Catalog: SDSS, {\it GALEX} and {\it WISE} data}

We used a spectroscopic sample of galaxies in 
  the SDSS Data Release 7 (SDSS DR7, Abazajian et al. 2009).
We complement these data with a photometric sample of SDSS galaxies
  whose redshift information is not available in the SDSS database,
  but available in the literature (Hwang et al. 2010).
  
We adopted the {\it GALEX} data for these SDSS galaxies
  from the {\it GALEX} GR6\footnote{http://galex.stsci.edu/GR6} that provides
  the cross-matched table (\textbf{xSDSSDR7}) against the SDSS DR7.
The matching tolerance is 5\arcsec ($\sim$FWHM of the {\it GALEX} PSF).
To avoid the contamination by nearby sources within the matching tolerance,
  we selected only unique matches:
  for a given SDSS galaxy, 
  we choose one {\it GALEX} object that is the closest to the SDSS object
  and vice versa.
We used only the sources covered by the Medium Imaging Survey (MIS)
  with a limiting magnitude of 22.7 (mag in AB) in the near-UV (Morrissey et al. 2005).
%  (MIS; exposure time range: 1000$-$5000 s).
Figure 1 shows the redshift distribution of the SDSS galaxies covered 
  by {\it GALEX} MIS.  
Roughly 5\% of the SDSS galaxies do not have {\it GALEX} counterparts; 
  these galaxies seem faint in the near-UV, and their NUV$-r$ colors are very red.
Some of them might be because of dust extinction and/or higher disk inclination.
To minimize the impact of those galaxies, we exclude galaxies with lower axis 
  ratio (i.e., $b/a$ less than 0.6) in the following analysis. 
%Thus, even among those galaxies detected at 12 $\mu$m, their mid-IR flux is as low as passively 
%  evolving old galaxies. 

%% Fig. 1  ----------------------------------------------------------------
\begin{figure}[h!]
\centering
\includegraphics[width=10cm]{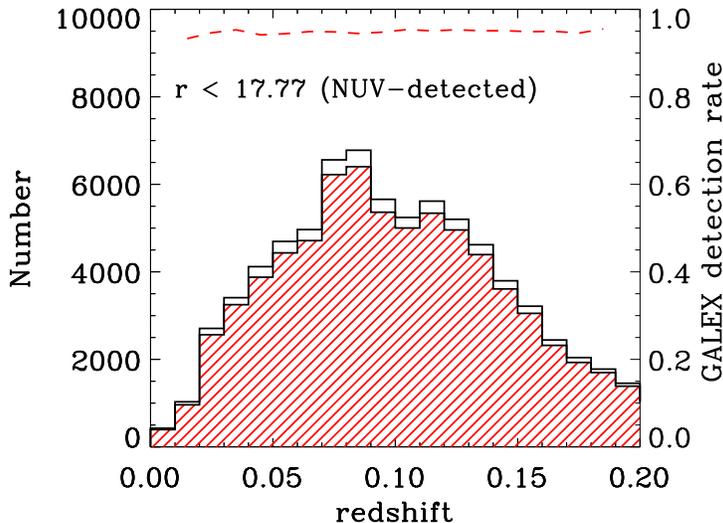}
\caption{Redshift distribution of SDSS galaxies covered by {\it GALEX} MIS. 
         Open and hatched histograms indicate all the galaxies regardless of 
         near-UV detection and the near-UV detected galaxies, respectively. 
         The dashed line is the ratio of the two histograms (i.e., {\it GALEX} 
         MIS near-UV detection rate).}
\end{figure}
%% ------------------------------------------------------------------ end Fig. 1

We used the mid-IR data obtained by the {\it WISE} satellite,
  which covers all the sky at four mid-IR bands (3.4, 4.6, 12 and 22 $\mu$m). 
The {\it WISE} all-sky source catalog\footnote{http://wise2.ipac.caltech.edu/docs/release/allsky/}
  contains photometric data for over 563 million objects.
We identified {\it WISE} counterparts of the SDSS galaxies
  by cross-correlating them with the sources in the {\it WISE} all-sky data release
  with a matching tolerance of 3\arcsec($\sim$ 0.5$\times$FWHM of the PSF at 3.4 $\mu$m).
We used the point source profile-fitting magnitudes,
 and restrict our analysis to the sources with 12 $\mu$m flux density, 
 $S_{12\mu m} \geq 0.7$ mJy where the {\it WISE} photometric completeness
 is larger than $\sim$90 \%.
{\it WISE} 5$\sigma$ photometric sensitivity is estimated to be better 
  than 0.08, 0.11, 1 and 6 mJy
  at 3.4, 4.6, 12 and 22 $\mu$m
  in unconfused regions on the ecliptic plane (Wright et al. 2010).
   
We also used several value-added galaxy catalogs (VAGCs) drawn from SDSS data.
We adopted the photometric parameters from the SDSS pipeline (Stoughton et al. 2002).
We used Petrosian magnitudes for $r$-band magnitudes of galaxies and model magnitudes
   for galaxy colors. 
We took the spectroscopic parameters including
   the 4000-\AA{} break $D_{n}$4000, the Balmer absorption-line index H$\delta_{A}$,
   and H$\alpha$ equivalent width from
%  SFRs (Brinchmann et al. 2004) and oxygen abundance (Tremonti et al. 2004) from 
  the MPA/JHU DR7 VAGC\footnote{http://www.mpa-garching.mpg.de/SDSS/DR7/}.
  We adopted galaxy morphology data and internal structure parameters 
  (concentration index $c_{in}$ in $i$-band and $g-i$ color gradient) from the
  Korea Institute for Advanced Study (KIAS) DR7 VAGC\footnote{http://astro.kias.re.kr/vagc/dr7/}
  (Choi et al. 2010).
In this catalog, galaxies are divided into two morphological types based on their locations 
  in the ($u-r$) color versus ($g-i$) color gradient space and in the $i$-band concentration 
  index space (Park \& Choi 2005): early (ellipticals and lenticulars) and late (spirals and
  irregulars) types. The resulting completeness and reliability for the morphological 
  classification reaches 90\%. We performed an additional visual check of the color images 
  of the galaxies misclassified by the automated scheme, and of the galaxies that are not 
  included in the KIAS DR7 VAGC. In this procedure, we revised the types of blended or merging
  galaxies, blue but elliptical-shaped galaxies, and dusty edge-on spirals.

For the following analysis, we used the galaxies at 0.04 $<$ z $<$ 0.11 and 
  $m_{r}$ $<$ 17.77 (magnitude limit for the SDSS main galaxy sample).
%To compare our results with previous studies in Section 3, we restrict galaxies with 
%  $M_{r}$ $<$ $-$21.5 and 0.04 $<$ z $<$ 0.11.
Figure 2 shows $r$-band absolute Petrosian magnitudes of 12 $\mu$m selected 
  galaxies with near-UV detection as a function of redshift (gray dots).
Red filled circles are spectroscopically selected quiescent galaxies on optical red sequence 
  (see the next section).
%We defined a volume-limited samples of galaxies with
%  $M_{r}$ $<$ $-$21.5 and 0.04 $<$ z $<$ 0.11 for the following analysis.
We often use a volume-limited samples of galaxies with
  $M_{r}$ $<$ $-$21.5 and 0.04 $<$ z $<$ 0.11 (solid lines).
At $z$ $<$ 0.04, SDSS spectroscopy is incomplete for bright galaxies due to the problem
  of small fixed-size aperture (Kewley et al. 2005).
The visual inspection of galaxy morphology becomes uncertain with increasing redshift, so
  $z$ $=$ 0.11 is our upper limit (Kaviraj et al. 2007; Schawinski et al. 2007).

%% Fig. 2  ----------------------------------------------------------------
\begin{figure}[h!]
%\resizebox{\hsize}{!}
\centering
\includegraphics[width=10cm]{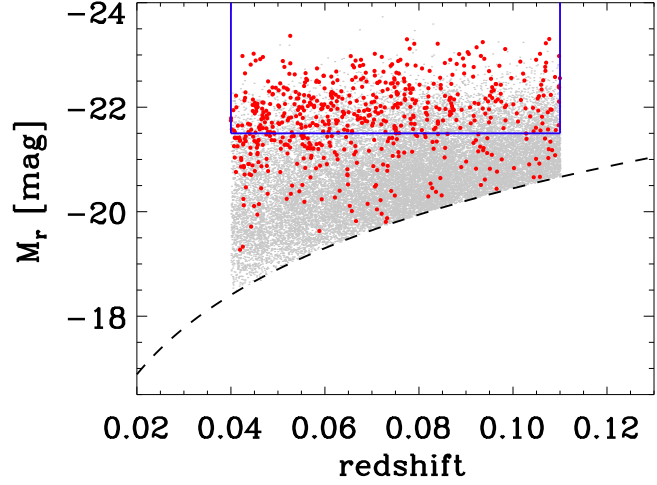}
\caption{Absolute $r$-band magnitude vs. redshift for {\it WISE} 12 $\mu$m selected galaxies with 
         {\it GALEX} near-UV detection. Red filled circles indicate spectroscopically selected 
         quiescent galaxies on the optical red sequence, and small dots indicate all the galaxies 
         regardless of optical colors. Solid lines define the volume limited sample 
         at 0.04 $<$ z $<$ 0.11 and $M_{r}$ $<$ $-$21.5.
         The bottom dashed curve indicates the apparent magnitude limit of 
         $m_{r}$ = 17.77 (Choi et al. 2007).}
\end{figure}
%% ------------------------------------------------------------------ end Fig. 2

In Figure 3, we plot the distributions of absolute magnitude $M_{r}$, 
 $u-r$ color, and NUV-$r$ color for the SDSS galaxies covered by {\it GALEX} MIS. 
The hatched and open histograms indicate the galaxies with 12 $\mu$m detection and those 
 regardless of 12 um detection, respectively.
The ratio of the two histograms represents the 12 $\mu$m detection rate and is overplotted as 
 a dashed line.
The 12 $\mu$m detection rate increases with $r$-band luminosity at $M_{r}$ $>$ $-$20.
However, it seems that the detection rate is nearly constant for galaxies at $M_{r}$ $<$ $-$20,
 which is mainly due to the bright, red galaxies with weak 12 $\mu$m emission. 
In terms of colors, the detection rate decreases sharply 
 at the red end because the red galaxies are likely to have weaker 12 $\mu$m emission than 
 the blue galaxies, irrespective of their luminosity. 
At the blue end, the detection rate is small because some fainter (less massive), blue galaxies 
 are not detected because of $r$-band magnitude limit.
Interestingly, the detection rate peaks at intermediate colors ($\sim$90\%),
 consistent with Salim et al. (2009).

%% Fig. 3  ----------------------------------------------------------------
\begin{figure}[h!]
\resizebox{\hsize}{!}
{\includegraphics{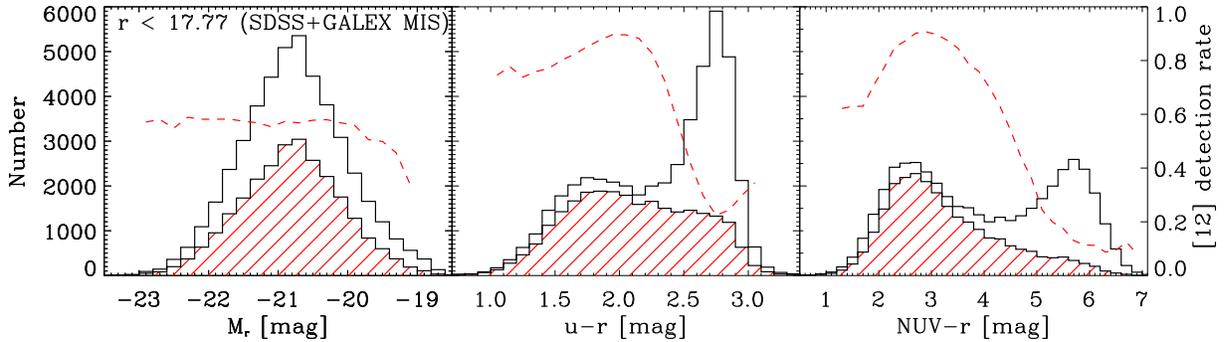}}
\caption{Histograms for $r$-band absolute magnitude $M_{r}$ ($left$),
        $u-r$ ($middle$), and  NUV-$r$ ($right$) for SDSS galaxies with GALEX near-UV detection
        and $r$ $<$ 17.77 (open histograms). Hatched histograms are for those of SDSS galaxies
        with near-UV and 12 $\mu$m detections. The dashed lines are the ratio of the two 
        histograms (i.e., 12 $\mu$m detection rate).}
\end{figure}
%% ------------------------------------------------------------------ end Fig. 3

\subsection{Sample: optically quiescent galaxies on the red sequence}

%To study the mid-IR and NUV emissions of quiescent red galaxies (hereafter quiescent red-sequence galaxies), 
We first used an an optical color-magnitude diagram to identify red-sequence galaxies among the
 12 $\mu$m emitters (see the top panel in Figure 4).
Among the 12 $\mu$m selected galaxies, 18\% (4549) are on the red sequence.
Among the 12 $\mu$m selected galaxies on the red sequence, 35\% are brighter than 
 $M_{r}$= $-$21.5; these galaxies will be often used for comparing with the results in other 
 studies. 
The top panel in Figure 4 shows the optical color-magnitude diagram for 12 $\mu$m selected 
 galaxies. 
The red-sequence is defined from a linear fit to the $u-r$ versus $M_{r}$, 
 by rejecting outliers iteratively based on the bi-weight calculation.
The standard deviation of residuals to the fit is 0.11 mag (1$\sigma$), indicating a 
tight red sequence. The horizontal solid line indicates the color cut
adopted in this study to separate red galaxies (redward of the solid line) from blue galaxies 
(blueward of the solid line). The color-magnitude relation is moved to a bluer color 
by $\Delta(u-r)=0.11 (1\sigma)$ to define the color cut. 
For comparison, we also plot the criterion determined in Choi et al. (2007, dotted line). 
Choi et al. used SDSS galaxies brighter than $M_{r}$ = $-$18.5, and rejected late-type 
 galaxies with axis ratios b/a $<$ 0.6 to reduce the bias caused by internal extinction. 
The $u-r$ colors of our red-sequence galaxies are slightly redder than their cut, 
 suggesting our color cut is more strict than previous studies. 
The slope of our red sequence is about $-$0.08, similar to that for luminous 
 red galaxies in SDSS (Baldry et al. 2004).
%Consequently, we select red galaxies as red as most bright early-type galaxies. Thus,
%the slope of our red sequence is about $-$0.08, which is consistent with Baldry et al. (2004) 
%who reported a slope for luminous red galaxies selected from the SDSS.

%Second, we adopted H$\alpha$ emission line cut of H$\alpha$ EW $>$ $-$1 \AA{} to select 
%spectroscopically quiescent galaxies on the red sequence. 
%We also excluded galaxies with $i$-band isophotal axis ratio $b/a$ less than 0.6 
%(see Choi et al. 2007) to minimize the impact of dust reddening due to inclination of galaxies, 
%since dusty SF galaxies tend to have weak emission lines highly obscured.

We then use H$\alpha$ equivalent width to identify the galaxies without current star formation. 
Among the red-sequence galaxies, we select the galaxies with H$\alpha$ equivalent width 
 $>$ $-$1 \AA{} as ``quiescent'' galaxies. We also excluded galaxies with $i$-band isophotal 
 axis ratio $b/a$ less than 0.6 (see Choi et al. 2007) to reject star-forming galaxies with large  
 dust obscuration.
These processes removed about 85\% of red-sequence galaxies, and left 
 red-sequence galaxies without current star formation.

%AGN removal
%Finally, AGNs are to be removed since they can produce UV and mid-IR flux.   
We also removed galaxies with active galactic nuclei (AGNs) that can produce UV and mid-IR emissions. 
As expected, most Type II AGNs are removed using the H$\alpha$ emission line cut.
%Furthermore, we determined the spectral types of galaxies using the criteria of Kewley ety al. (2006)
%based on the BPT analysis of Baldwin et al. (1981). 
we removed the remaining Type II AGNs (i.e., Seyferts, LINERs, and composites) using the spectral
 types determined by the criteria of Kewley et al. (2006) based on the emission-line ratio 
 diagrams (Baldwin et al. 1981).
We also removed Type I AGNs with broad Balmer lines with quasar spectral classifications 
provided by the SDSS pipeline (i.e., \texttt{specClass} = \texttt{SPEC}$\_$\texttt{QSO} or \texttt{SPEC}$\_$\texttt{HIZ}$\_$\texttt{QSO}; see Stoughton et al. 2002  for more details).
However, there could be still unidentified AGNs in our sample. 
There are some galaxies without SDSS spectra because their redshifts are from the literature.
In some galaxies, the AGN signature could be hidden by dust (e.g., Lee et al. 2012). 
To identify these dusty AGNs, we used the {\it WISE} color-color selection criteria proposed by 
Jarrett et al. (2011). 
In summary, we identified 12 AGNs based on the optical spectra (i.e., Type I or II AGNs) and 
no AGNs based on the {\it WISE} color distribution.
We rejected these AGNs from the galaxy sample. 

In the result, we obtained 648 optically quiescent red-sequence galaxies
 \footnote{Hereafter we refer to those galaxies as ``quiescent red-sequence" galaxies.} 
 without H$\alpha$ emission and the activity in galactic nuclei for the following analysis.
Among them, 89\% and 11\% are morphologically early- and late-type galaxies, respectively.

%We found that the 12 $\mu$m selected red-sequence galaxies are composed of a 
%mixture of 5\% SF, 29\% AGN, 60\% undetermined and 6\% ambiguous galaxies. 

%% Fig. 4  ----------------------------------------------------------------
\begin{figure}[h!]
\centering
\includegraphics[width=12cm]{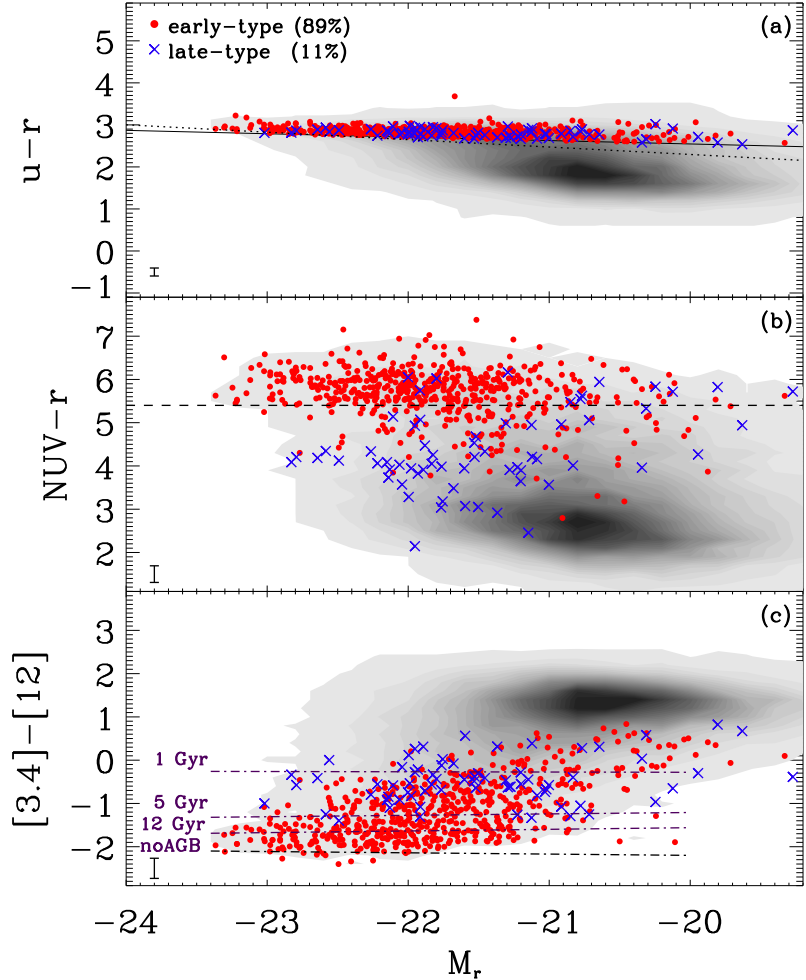}
\caption{Optical ($top$), near-UV ($middle$) and mid-IR ($bottom$) color-magnitude diagrams for
         quiescent red-sequence galaxies. 
         Red filled circles and blue crosses show early- and late-type galaxies, respectively. 
         Grayscale map indicates the galaxy number density for 12 $\mu$m selected galaxies. 
         Solid line indicates the color cut used in this study to identify red galaxies (redward 
         of the solid line). The color deviation from the color-magnitude relation is defined as 
         $\Delta(u-r)=0.11 (\sim1\sigma)$, where $\sigma$ is the standard deviation of 
         residuals to the color-magnitude relation fit. 
         The dotted line indicates the color cut used in Choi et al. (2007) 
         to divide the galaxies into red and blue galaxies. The dashed line in (b) indicates 
         the cutoff (NUV$-r$ = 5.4) for recent star formation in Schawinski et al. (2007). 
         The dot-dashed lines in (c) indicate model predictions calculated from the P03 AGB model 
         SSPs, assuming a metallicity sequence at three different stellar ages (1, 5 , and 12 Gyr), 
         respectively. The bottom line (`noAGB') represents the P03 model SSPs without AGB dust. 
         The error bars in the lower left corners show median color errors.}
\end{figure}
%% ------------------------------------------------------------------ end Fig. 4

\subsection{Subsamples}

Although we exclude galaxies with H$\alpha$ emissions and AGNs, quiescent red-sequence 
 galaxies still show a wide range of colors in the near-UV and mid-IR (Figure 4(b,c)). 
This suggests that the optical red-sequence consists of various types of galaxies.
We thus classify quiescent red-sequence galaxies into three subclasses based on NUV$-r$ and  
 [3.4]$-$[12] colors.

(1) quiescent red-sequence galaxies with near-UV excess: NUV$-r$ $<$ 5.4.
These galaxies have recent star formation (Schawinski et al. 2007).
Example optical color images of these galaxies are shown in the top panel in Figure 5.

(2) quiescent red-sequence galaxies with mid-IR excess: [3.4]$-$[12] $>$ $-$1.3. 
These galaxies have mid-IR excess emission over the stellar component (Ko et al. 2012). 
Example color images of these galaxies are in the middle panel in Figure 5.

(3) quiescent red-sequence galaxies without near-UV and mid-IR excess: NUV$-r$ $>$ 5.4 and 
 [3.4]$-$[12] $<$ $-$1.3.
These galaxies are completely quiescent in the optical, the near-UV and the mid-IR. 
Example color images of these galaxies are in the bottom panel in Figure 5.

The color images of the galaxies in three subclasses indicate that they all look like typical 
 bulge-dominated systems with red colors and good symmetry regardless of their NUV$-r$ and 
 [3.4]$-$[12] colors (see Figure 5).

%\subsubsection{Quiescent red-sequence galaxies with NUV excess}

%\textbf{We adopt the NUV$-r$ color cut used for selecting galaxies with recent star formation
%(Schawinski et al. 2007): NUV$-r$ $<$ 5.4 (i.e., NUV excess). Typical image of this class is shown in Figure 5(a).}

%\subsubsection{Quiescent red-sequence galaxies with mid-IR excess}

%\textbf{We adopt the [3.4]$-$[12] color cut (Ko et al. 2012) to consider galaxies to have 
% mid-IR excess emission over stellar emission:  [3.4]$-$[12] $>$ $-$1.3 (i.e., mid-IR excess). 
% Typical image of this class is shown in Figure 5(b).}

%\subsubsection{Quiescent red-sequence galaxies without NUV and mid-IR excess}

%\textbf{This class is completely quiescent (i.e., not only in optical but also in the near-UV and mid-IR). Typical image of this class is shown in Figure 5(c).}

%% Fig. 5  ----------------------------------------------------------------
\begin{figure}
\centering
\includegraphics[width=10cm]{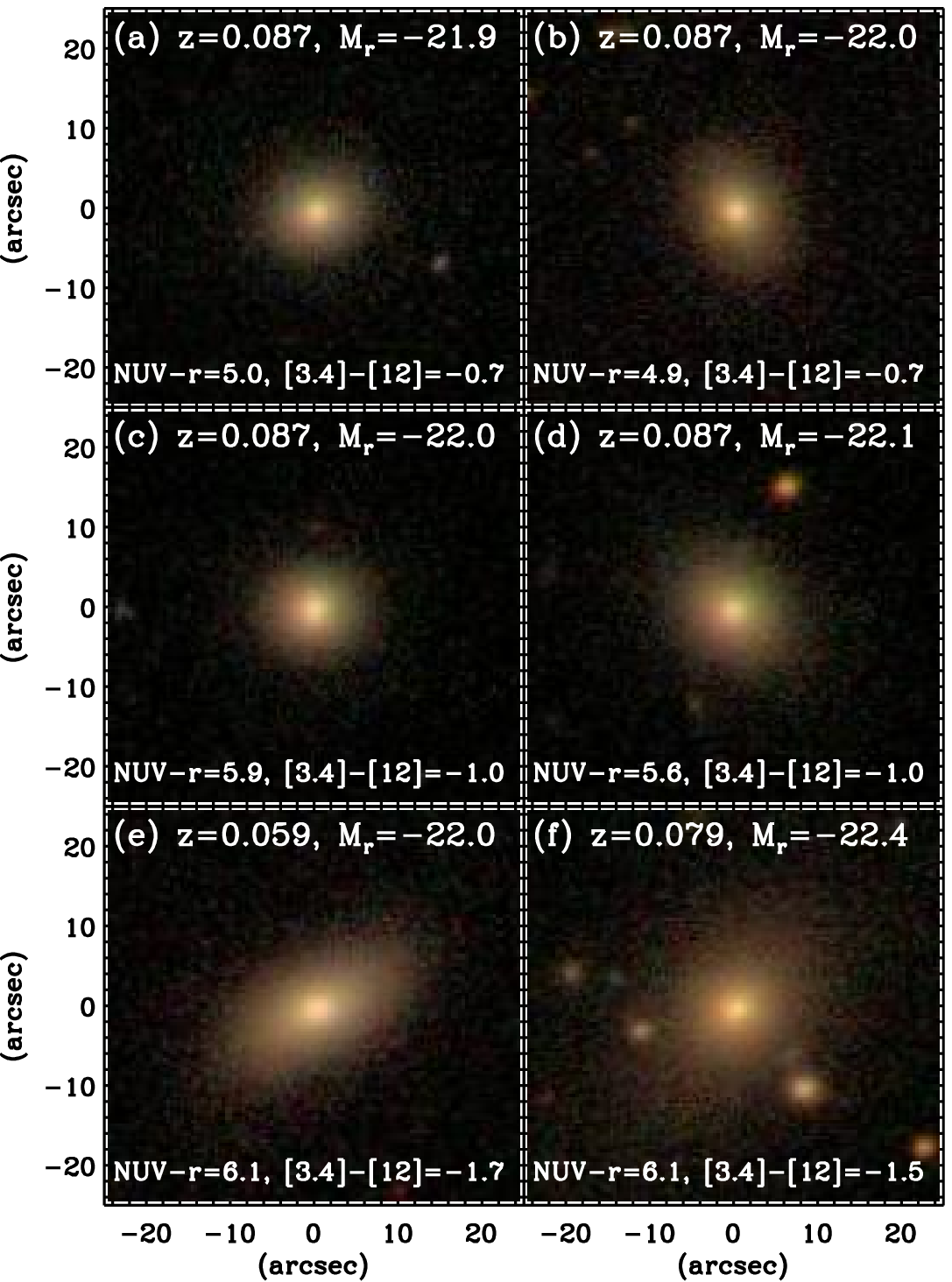}
\caption{Example optical color images of quiescent red-sequence galaxies. 
         Each image lists redshift, $r$-band 
         absolute magnitude $M_{r}$, NUV$-r$ and [3.4]$-$[12] colors. 
         Each low shows two objects in each subclass (see Section 2.3): 
         quiescent red-sequence galaxies with near-UV excess ((a) SDSS ObjID: 587725040090742995
         and (b) ObjID: 587724197209899203), those with mid-IR excess 
         ((c) ObjID: 587725041170972858 and (d) ObjID: 587725041171038336), and 
         those without near-UV and mid-IR excess ((e) ObjID: 587724241761796273 and 
         (f) ObjID: 587724233174876337).}
\end{figure}
%% ------------------------------------------------------------------ end Fig. 5

\section{RESULTS AND DISCUSSION}

\subsection{near-UV and mid-IR Excess Emissions of Quiescent Red-sequence Galaxies}

\subsubsection{near-UV and mid-IR Color-Magnitude Diagrams}

In Figure 4(b), we show the NUV$-r$ color-magnitude diagram for quiescent red-sequence galaxies. 
%The dashed line represents the NUV$-r$ cut used for selecting galaxies with RSF
%(Schawinski et al. 2007): NUV$-r$ $<$ 5.4 (i.e., NUV excess).
The dashed line indicates the NUV$-r$ color cut used for selecting galaxies with near-UV excess 
(i.e., NUV$-r$ $<$ 5.4).
This figure shows that 26\% of quiescent red-sequence galaxies have NUV$-r$ $<$ 5.4,
 indicating their recent star formation.
The fraction decreases to 20\% for bright ($M_{r}$ $<$ $-$21.5), quiescent red-sequence galaxies.
If we consider only early-type, bright, quiescent red-sequence galaxies, the fraction is 
14 $\pm$ 2\%. 
This is roughly a half of the result in Schawinski et al. (2007) who 
 found that 30\% $\pm$ 2\% of early-type galaxies with $M_{r}$ $<$ $-$21.5 show recent star formation. 
The reason for the difference between the two studies is mainly because of different 
 sample selection; we use only bright ($M_{r}$ $<$ $-$21.5) early-type galaxies 
 on the red sequence at $m_{r}$ $<$ 17.77 by rejecting galaxies with axis ratio 
 $b/a$ $<$ 0.6, but Schawinski et al. (2007) use early-type galaxies at $m_{r}$ $<$ 16.8 
 regardless of axis ratio. 
%Some of the difference might also be due to our morphological classification is not 
% the same as theirs.

Interestingly, 74\% (52 out of 70) of late-type, quiescent red-sequence galaxies show 
 near-UV excess emission, much larger than the fraction for early-type galaxies 
 (i.e., 20\%, 116 out of 578). 
%Therefore, the NUV-excess fraction of late types is much higher than early types by more 
% than three times. 
The difference is larger for bright galaxies with $M_{r}$ $<$ $-$21.5.
These results suggest that the signs of recent star formation in galaxies are closely related 
 to galaxy morphology and optical luminosity. 
Some of late-type, quiescent red-sequence galaxies with near-UV excess could be disk galaxies 
 whose star formation have been gradually quenched (e.g., Gebhardt et al. 2003) 
 or suddenly stopped by external mechanisms including galaxy mergers within $\sim$1 Gyr.
%But, separation between two explanations is beyond the scope of this paper. 

In Figure 4(c), we show the {\it WISE} [3.4]$-$[12] versus $M_{r}$ color-magnitude diagram for 
 quiescent red-sequence galaxies. 
We also overplot the predictions from Single Stellar Population (SSP) models that include the dust
 emission from circumstellar dust around AGB stars (dot-dashed lines; Piovan et al. 2003; 
 hereafter P03).
The optical color-magnitude relation can be well described by a single age model with a 
 metallicity gradient. 
The same model fits the [3.4]$-$[12] versus $M_{r}$ color-magnitude diagram at three different 
 stellar ages (1, 5, and 12 Gyr), shown by dot-dashed lines. 
The bottom line represents the SSP model without AGB dust. 
Interestingly, any single age model fails to reproduce the dispersion in the [3.4]$-$[12] colors 
 (Ko et al. 2009, 2012), which suggests either the existence of younger stellar populations 
 or some other mechanisms.
%Although the circumstellar dust formation and the evolution of AGB stars are still known to be
%uncertain in detail, the excess emission in 12 $\mu$m is certain. 
The existence of excess emission at 12 $\mu$m seems robust even though the detailed physics about 
 the circumstellar dust formation and the evolution of AGB stars responsible for the excess 
 emission is not understood completely.
We then can set a threshold in mid-IR color ([3.4]$-$[12]) to consider galaxies with
 mid-IR excess emission over stellar emission (mid-IR excess). 
We adopt the cut of [3.4]$-$[12] = $-$1.3 (Ko et al. 2012). 
Galaxies redder than this cut are considered to be relatively young ($<$ 5 Gyr) quiescent 
 red-sequence galaxies. 
Among our quiescent red-sequence galaxies, 55\% show mid-IR excess emission. 
This fraction which much higher than the for galaxies with near-UV excess emission. 
Interestingly, nearly all (67 out of 70) late-type, quiescent red-sequence galaxies show 
 mid-IR excess emission, much larger than the fraction for early-type, quiescent red-sequence  
 galaxies (51\%, 292 out of 578).
%Therefore, the mid-IR-excess fraction for late types are roughly two times higher than early types.
If we consider bright ($M_{r}$ $<$ $-$21.5), early-type, quiescent red-sequence galaxies, 
 the fraction would be 39\% (156 out of 397). 
This is comparable to the result of Ko et al. (2012; see their Figure 25); they found that 
 $\sim$42\% of massive ($>$ 10$^{10}$ $M_{\odot}$), early-type, red-sequence galaxies in 
 the outskirts of galaxy clusters show mid-IR excess emission.
%Furthermore, Ko et al. (2012) showed that the fraction of mid-IR-excess decreases as the galaxy 
%number density increases and in the cluster cores.

\subsubsection{near-UV and mid-IR Color-Color Diagram}

%To probe the origin of NUV- and mid-IR-excess emission in QRGs, we performed a two-component SSPs analysis 
%by the mixing of young and intermediate-age stars with an old (12 Gyr) underlying component.

In Figure 6(a,b), we show the distributions of mid-IR and near-UV colors for quiescent 
 red-sequence galaxies with $M_{r}$ $<$ $-$21.5 (hatched histograms). 
We use this magnitude limit to select the galaxies not affected by the 12 $\mu$m detection limit
 (see Figure 4(c)). For comparison, we also plot the distributions of all the galaxies at 
 $M_{r}$ $<$ $-$21.5 and 0.04 $<$ $z$ $<$ 0.11 with open histograms.
%We restrict {\it WISE} sources with S/N $>$ 3 at 3.4 and 12 $\mu$m.  
The hatched histogram in the top panel (a) shows a peak around the quiescent, old galaxies, 
and have a red tail consisting of galaxies with mid-IR excess. The hatched histogram in the right 
panel (b) also shows a peak around the quiescent, old galaxies, but have a blue tail consisting 
of galaxies with near-UV excess. We overplot the eye-fitted Gaussians to the histograms of quiescent, old galaxies, which give $\sigma$ values of 0.26 for [3.4]$-$[12] colors and of 0.35 for NUV$-r$ colors. These $\sigma$ values roughly correspond to the maximum errors of colors. The cross in the lower right corner in (c) indicates median errors.

Figure 6(c) shows the [3.4]$-$[12] vs. NUV$-r$ color-color distribution of 436 
 bright ($M_{r}$ $<$ $-$21.5), quiescent red-sequence galaxies. 
Among them, 35 galaxies are not detected in the near-UV.
%No galaxies that are not detected in the near-UV have NUV excess if consider the {\it GALEX} MIS 
%limiting magnitude (22.7 in the near-UV). 
Considering the {\it GALEX} MIS limiting magnitude (22.7 mag in the near-UV), these galaxies not 
 detected in the near-UV do not have near-UV excess emission.
However, 14 galaxies of them have mid-IR excess emission.
The vertical and horizontal dashed lines represent the 
 mid-IR excess cut of [3.4]$-$[12] = $-$1.3 and the near-UV excess cut of NUV$-r$ = 5.4,  
 respectively. 
%The dot-dashed line indicates [3.4]$-$[12] = 0 cut used in Ko et al. (2012) to separate 
%SF galaxies from quiescent ones.
% when the IR luminosity is considered as a SF indicator. 
Interestingly, most quiescent red-sequence galaxies with near-UV excess show mid-IR excess 
 emission, but only 36\% of quiescent red-sequence galaxies with mid-IR excess show near-UV excess 
 emission. 
Among these bright ($M_{r}$ $<$ $-$21.5), quiescent red-sequence galaxies, the fraction of 
 galaxies with mid-IR excess is about 44\%, roughly twice higher than the fraction of galaxies 
 with near-UV excess ($\sim$20\%).
%almost all NUV-excess QRGs show mid-IR-excess emission, while only 36\% 
%of mid-IR-excess QRGs show NUV-excess emission. Therefore, among bright QRGs, $\sim$44\% have mid-IR excess, 
%which is roughly two times higher than the NUV-excess fraction ($\sim$20\%). 
%Interestingly, among QRGs early-type fraction of quiescent (i.e. QRGs without NUV- and mid-IR-excess), 
%only mid-IR-excess 
%(i.e. QRGs with [3.4]$-$[12] $>$ $-$1.3 and NUV$-r$ $>$ 5.4), and NUV-excess are $\sim$100\%, 96\%, 
%and 67\%, respectively. In other words, about 9\% QRGs have late-type morphologies and they show 
%NUV-excess emission (i.e. recent star formation within 1 Gyr).
%This late-type QRGs showing NUV-excess are likely to be explaned as spirals that are gradually 
%quenching their star formation, instead of QRGs showing recent star formation within 1 Gyr. But,
%separation between two explanations is beyound the scope of this paper. 

To study the physical origin of these near-UV and mid-IR excess emissions of quiescent 
 red-sequence galaxies, we perform a two-component SSP analysis.
% by the mixing of young and intermediate-age stars with an old (12 Gyr) underlying component.
We overplot the two-component SSP model grids that are the combination of an old (12 Gyr) 
 underlying population with young (0.5, 1, and 2 Gyr) populations with a solar metallicity.
We vary the fraction of young component to the old one. 
Having only $\sim$1\% of 0.5 Gyr and $\sim$5\% of 1 Gyr populations can make the NUV$-r$ colors of 
 galaxies be in the near-UV excess region, consistent with previous results of near-UV studies of 
 early-type galaxies (e.g., Yi et al. 2005; Kaviraj et al. 2007).
Similarly, the small amount of young (0.5 and 1 Gyr) populations also make galaxies have 
 mid-IR excess emission. 
The NUV$-r$ color, however, is insensitive to 2 Gyr population, 
 while the [3.4]$-$[12] color is still sensitive to an intermediate-age (2 Gyr) population. 
This means that although both near-UV and mid-IR are sensitive to very recent ($<$ 1 Gyr) star 
 formation, only the mid-IR is sensitive to star formation over longer (up to $\sim$2 Gyr) 
 timescales.
 
For comparison, we also show 12 $\mu$m selected E+A galaxies from Choi et al. (2009) in Figure 
 6(c). 
E+As are post-starburst (within $\sim$1 Gyr) systems with strong Balmer absorption lines
 and weak H$\alpha$ emission line. 
The figure shows that most E+As have near-UV and mid-IR excess emissions except two galaxies that 
 have only mid-IR excess emission. 
%The colors of E+As are also expected to be consistent with $\sim$1 Gyr population models. 
As expected, the near-UV and mid-IR colors of E+As are consistent with predictions of $\leq$ 1 Gyr 
 models with $>$ 10\% of young component. 
This amount of young component for E+As is larger than that for quiescent red-sequence galaxies  
 with near-UV excess. 
However, it should be noted that E+As are not red-sequence galaxies, so they  
 are not in the same evolutionary stage as quiescent red-sequence galaxies.
   
We also show the SWIRE templates of Polletta et al. (2007) for comparison: 
 3 ellipticals (2, 5, 13 Gyr), 7 spirals (S0, Sa, Sb, Sc, Sd, Sdm, Spi4), 6 starbursts 
 (M82, Arp220, N6090, N6240, I20551, I22491), and 2 AGNs (QSO1 and QSO2). 
These are generated with the GRASIL code (Silva et al. 1998) including dusty envelopes of AGB 
 stars following the prescription by Bressan et al. (1998). 
In particular, the mid-IR spectra of spiral and starburst templates at 5$-$12 $\mu$m are adopted 
 from various observed spectra of galaxies (Polleta et al. 2007). 
Although the GRASIL model predicts the dust emission from evolved stars differently from P03 
 models, the expected near-UV and mid-IR colors for early-type (ellipticals, S0 and Sa) galaxies 
 are broadly consistent with our data (i.e., the [3.4]$-$[12] color is redder as mid-IR 
 weighted mean stellar age decreases).
%Although the GRASIL model have difference in how dust emission from evolved stars is predicted,
%their early-type (ellipticals, S0 and Sa) galaxies are positioned broadly in agreement that 
%the [3.4]$-$[12] color shows redder as the mean stellar age decreases.

%% Fig. 6  ----------------------------------------------------------------
\begin{figure}
\centering
\includegraphics[width=12cm]{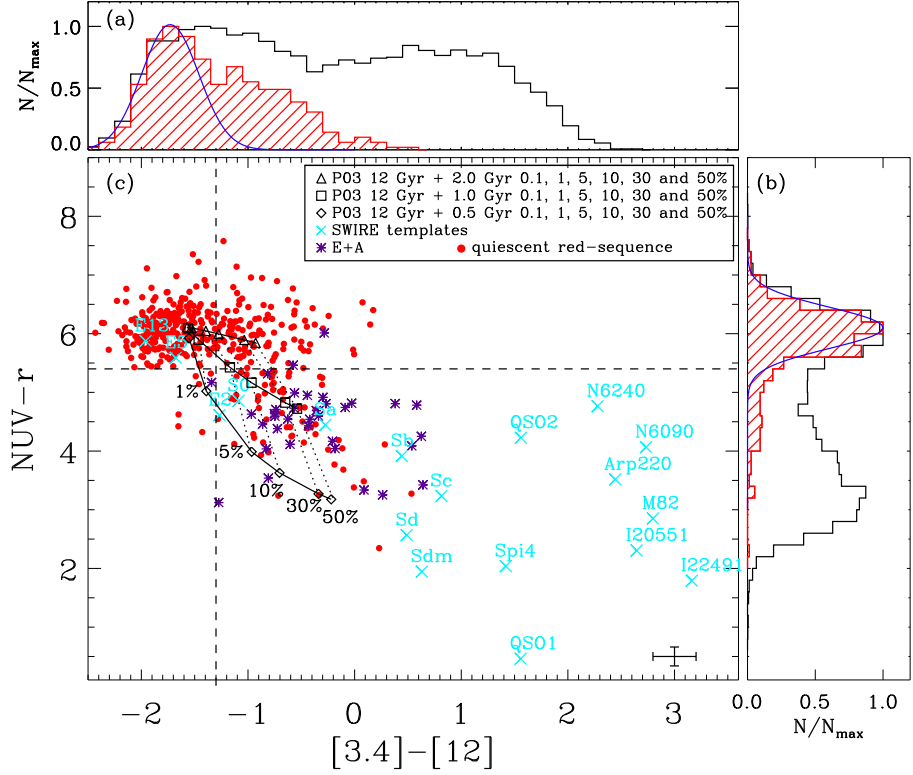}
\caption{Histograms of mid-IR (a) and near-UV (b) colors for bright ($M_{r}$ $<$ $-$21.5), quiescent
         red-sequence galaxies. Open histograms show all the galaxies 
         with $M_{r}$ $<$ $-$21.5 and 0.04 $<$ $z$ $<$ 0.11 regardless of optical colors,
         H$\alpha$ equivalent width, and axis ratio.
         Eye-fitted Gaussians are traced over the histograms with $\sigma$ values of 0.26 
         (for [3.4]$-$[12] color) and 0.35 (for NUV$-r$ color), which roughly corresponds 
         to the maximum errors of colors. For comparison, the cross in the lower right corner
         indicates median errors.
         Two histograms have the main peaks of quiescent, old galaxies, and the red tail (a) 
         of quiescent red-sequence galaxies with mid-IR excess while the blue tail (b) of 
         quiescent red-sequence galaxies with near-UV excess.
         [3.4]$-$[12] vs. NUV$-r$ color-color distribution of bright, 
         quiescent red-sequence galaxies (c). We show the SWIRE templates of Polletta et al. (2007) including  3 ellipticals 
         (2, 5, 13 Gyr), 7 spirals (S0, Sa, Sb, Sc, Sd, Sdm, Spi4), 6 starbursts 
         (M82, Arp220, N6090, N6240, I20551, I22491), and 2 AGNs (QSO1 and QSO2). 
         We also plot the two-component SSP model grid (P03). 
         We combine an old (12 Gyr) underlying population with young (0.5, 1, and 2 Gyr) populations
         of a solar metallicity, with varying the fraction of young component to the old one.
         The vertical and horizontal dashed lines represent the mid-IR excess cut of 
         [3.4]$-$[12] = $-$1.3 and the near-UV excess cut of NUV$-r$ = 5.4, respectively. 
         For comparison, asterisks represent E+A galaxies from Choi et al. (2009).}
\end{figure}
%% ------------------------------------------------------------------ end Fig. 6

\subsubsection{near-UV and mid-IR emissions from young starburst galaxies?}

Although we remove galaxies with current star formation using H$\alpha$ emission and axis ratios
 (see Section 2.2), there could be a still contribution from young starburst populations to 
 near-UV or mid-IR excess emissions. For example, the localized thin dust absorption can change
 NUV$-r$ colors of galaxies, and the variation in the 12 $\mu$m emission around deeply embedded 
 sources can change [3.4]$-$[12] colors of galaxies (see Figure 6).

To examine this possibility, we first checked how many quiescent red-sequence galaxies 
 are detected at 22 $\mu$m. 
Among 436 bright, quiescent red-sequence galaxies, 385 ($\sim$88\%) galaxies are not
 detected at 22 $\mu$m (i.e., signal-to-noise ratio at 22 $\mu$m $<$ 3), indicating that 
 most of them are not dusty, star-forming systems.
For the remaining 51 galaxies, we plot [3.4]$-$[12] colors versus [3.4]$-$[22] colors 
 in Figure 7 to examine how their star formation activity is different from other galaxy  
 populations.
To do that, we also plot the median value of blue, star-forming galaxies at 
 0.04 $<$ $z$ $<$ 0.11 (open square). 
%Because 22 and 3.4 um data are good proxies for star formation rates (Jarrett et al. 2013, ApJ, %145, 6) and stellar masses (Hwang et al. 2012, ApJ, 758, 25), respectively, [3.4]-[22] colors can %be a good indicator of specific star formation rate.
The figure shows that the [3.4]$-$[22] colors of quiescent red-sequence galaxies with 
 mid-IR excess (blue filled circles) is, on average, bluer than for blue, star-forming galaxies. 
Half of these galaxies with mid-IR excess have [3.4]$-$[22] colors as blue as quiescent 
 red-sequence galaxies without mid-IR excess (i.e., quiescent, old galaxies; red filled circles). 
%These suggest that quiescent red-sequence galaxies with mid-IR excess have a lack of 
% 22 $\mu$m emission, compared to blue star-forming galaxies.
These suggest that although some quiescent, red-sequence galaxies with mid-IR excess are detected  
 at 22 $\mu$m, their (recent) star formation activity is not as strong as blue, star-forming galaxies.

In summary, quiescent red-sequence galaxies can have near-UV and/or mid-IR excess emissions 
 because of recent star formation without ongoing star formation. 
Their optical colors are red mainly because of their underlying old stellar populations.
%  (not an effect of dust reddening)
This can suggest that they have just arrived to the red sequence after recent star formation 
 activity. 
Among the quiescent red-sequence galaxies, there could be currently weakly 
 star-forming galaxies, but the fraction of these galaxies is very low.

%However, we caution that the mid-IR emission may arise from low level of dusty star-forming %galaxies. In our sample, their fraction roughly reaches to $\sim$21\%.} 
%Furthermore, it is likely that late-type QRGs are relatively younger than early-type QRGs.

%% Fig. 7  ----------------------------------------------------------------
\begin{figure}
\centering
\includegraphics[width=12cm]{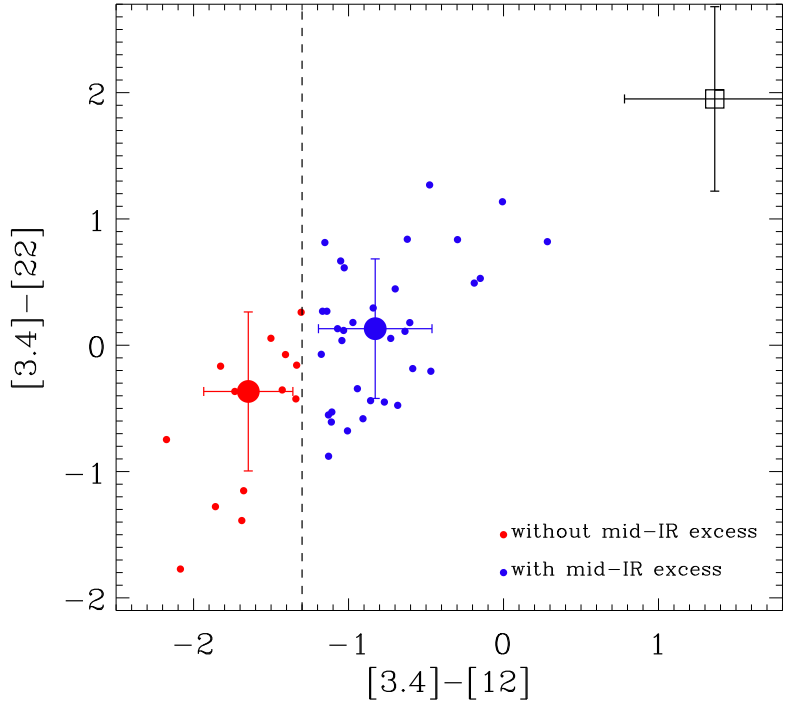}
\caption{[3.4]$-$[12] vs. [3.4]$-$[22] colors for $\textit{WISE}$ 22 $\mu$m detected 
         (i.e., signal-to-noise ratio at 22 $\mu$m $\geq$ 3) 
         quiescent red-sequence galaxies with $M_{r}$ $<$ $-$21.5. Large red and blue  
         circles are median values of colors for quiescent red-sequence
         galaxies without and with mid-IR excess, respectively. 
         For comparison, we plot the median value of blue, star-forming galaxies at 
         0.04 $<$ $z$ $<$ 0.11 (open square).
         The vertical dashed line indicates [3.4]$-$[12] = $-$1.3 cut used in this study 
         for selecting quiescent red-sequence galaxies with mid-IR excess.}
\end{figure}
%% ------------------------------------------------------------------ end Fig. 7

\subsection{Comparison of Physical Properties between mid-IR and near-UV Excess Quiescent Red-sequence Galaxies}

%To study the recent star formation history of bright ($M_{r}$ $<$ $-$21.5) ETGs with NUV and 
%mid-IR-excess emissions, 
%we divide quiescent red-sequence galaxies into three classes depending on their NUV$-r$ and %[3.4]$-$[12] colors in Section 2.3. 
% 1) `pure QRGs' that are ETGs without NUV- and mid-IR-excess emissions (i.e., [3.4]$-$[12] %$<$ $-$1.3 
% and NUV$-r$ $>$ 5.4, hereafter p-QRGs), 
% 2) `only mid-IR-excess QRGs' that are ETGs showing mid-IR-excess emission but no NUV-excess %emission 
% (i.e., [3.4]$-$[12] $>$ $-$1.3 and NUV$-r$ $>$ 5.4, hereafter mid-IR-QRGs), and 
%3) `NUV-excess QRGs' that are ETGs showing NUV-excess emission (NUV$-r$ $<$ 5.4, hereafter NUV-QRGs).
%It should be noted that we restrict our analysis in this section to early-type galaxies.

In Figure 8, we show the environmental dependence of optical properties for three classes 
 of quiescent red-sequence galaxies: quiescent red-sequence galaxies with near-UV 
 excess, those with mid-IR excess (but no near-UV excess), and those without near-UV 
 and mid-IR excess.
%Although most ($\sim$91\%) of quiescent red-sequence are early-type galaxies, we exclude late-type galaxies to 
%study the recent star formation history of bright, red early-type galaxies with near-UV and mid-IR excess emissions. 
we exclude late-type galaxies in this plot to focus on the recent star formation history of 
 bright, red \textit{early-type} galaxies with near-UV and mid-IR excess emissions. 
 
As an environmental indicator, we use a surface galaxy number density, 
 $\Sigma_{5}$ that is defined by $\Sigma_{5}$ = 5($\pi D_{p,5}^{2}$)$^{-1}$. $D_{p,5}$ is 
 the projected distance to the 5th-nearest neighbor. 
The 5th-nearest neighbor of each target galaxy was identified among the neighbor galaxies  
 with $M_{r}$ $\leq$ $-$20.66 (equivalent to $r$=17.77 at $z$ = 0.11, see Figure 2) 
 that have velocities relative to the target galaxy less than 1500 km s$^{-1}$.
$\Sigma_{5}$ is a useful indicator for local density (e.g., Baldry et al. 2006; Muldrew et al. 
 2012).
It probes different scales depending on tracers; in this study,
 $\Sigma_{5}$  probes the physical scale of $\sim$0.5-10 Mpc (see Figure 8). 
If we use other environmental indicators (e.g., surface galaxy number density within a 
 projected distance of 0.5 or 1.0 Mpc; Ko et al. 2012), our conclusion does not change.

The top panel of Figure 8 shows $r$-band absolute magnitudes of quiescent red-sequence
galaxies as a function of local density. 
We also show E+A galaxies in Choi et al. (2009) for comparison (open stars). 
The quiescent red-sequence galaxies with recent star formation (red and blue filled stars
 for mid-IR and near-UV excess, respectively) seem fainter (or less massive)
 than those without recent star formation (solid line) at nearly all density range. 
This difference is more significant in high and intermediate density regions. 
Interestingly, the $r$-band magnitudes of galaxies with mid-IR and near-UV excess do not  
 change much with local density, but those of galaxies without mid-IR and near-UV excess show 
 a dependence on the local density.

In the middle panel of Figure 8, we plot the strength of the 4000-\AA{} break $D_{n}$4000 
(good indicator of the mean stellar age of a galaxy) and the Balmer absorption-line index
H$\delta_{A}$ (indicator of the recent star formation activity of a galaxy; 
Kauffmann et al. 2003). 
The figure shows that the median values of $D_{n}$4000 for quiescent red-sequence galaxies
with mid-IR and near-UV excess are similar to those without mid-IR and near-UV excess, independently
of environment. 
There is a hint of slightly lower $D_{n}$4000 of those with mid-IR and near-UV excess 
than those without mid-IR and near-UV excess (i.e., younger age of those with mid-IR and near-UV
excess than those without mid-IR and near-UV excess), but the difference is small. 
However, this age difference is clearly apparent in Figure 6(c); the mid-IR weighted mean
stellar ages of quiescent red-sequence galaxies with mid-IR excess are larger than 
those with near-UV excess, and smaller than those without mid-IR and near-UV excess.
%Although it seems that they have slightly lower $D_{n}$4000 than those of p-QRGs
% (indicating mid-IR- and NUV-QRGs are slightly younger than p-QRGs), this is just below 1 $\sigma$.   
Interestingly, the median values of H$\delta_{A}$ for quiescent red-sequence galaxies with
mid-IR and near-UV excess are also comparable to those without mid-IR and near-UV excess.
This suggests that Balmer lines are not sensitive to detect quiescent red-sequence 
 galaxies with mid-IR and near-UV excess emissions.
Because of definition of E+A galaxies, they have higher H$\delta_{A}$ values than quiescent 
 red-sequence galaxies. 
However, their near-UV colors are consistent with those of quiescent red-sequence galaxies 
 with near-UV excess (see Figure 6(c)). 
These suggest that quiescent red-sequence galaxies with near-UV excess might have 
 experienced weaker starbursts (or shorter starburst duration) than E+A
 galaxies within $\sim$1 Gyr (Choi et al. 2009). 
On the other hand, quiescent red-sequence galaxies with mid-IR excess might have experienced 
 recent star formation $\gtrsim$1 Gyr ago, so the contribution of A-type stars to their
 spectra is small.

We show the $i$-band concentration index ($c_{in}$) and ($g-i$) color gradient of quiescent red-sequence galaxies in the bottom panel of Figure 8.
%We adopt the values of ($g-i$) color gradient, concentration index ($c_{in}$), and Petrosian 
 %radius $R_{Pet}$ in KIAS VAGC (Choi et al. 2010). 
The ($g-i$) color gradient was defined by the color difference between the region with 
 $R$ $<$ 0.5$R_{Pet}$ and the annulus with 0.5$R_{Pet}$ $<$ $R$ $<$ $R_{Pet}$, 
 where $R_{Pet}$ is the Petrosian radius measured in $i$-band image. 
To account for the effect of flattening or inclination of galaxies, elliptical annuli 
 were used to calculate the parameters (Choi et al. 2007). 
The (inverse) concentration index is defined by $R_{50}/R_{90}$, where $R_{50}$ and $R_{90}$ 
 are semimajor axis lengths of ellipses containing 50\% and 90\% of the Petrosian flux in the 
 $i$-band image, respectively.
 
The figure shows that all the three classes of quiescent red-sequence galaxies have similar 
concentration index and ($g-i$) color gradients regardless of local density. 
%However, the color gradients of all the QRGs are on average lower in low-density regions than 
%in high-density regions. 
On the other hand, there are relatively many quiescent red-sequence galaxies with low color 
gradients (i.e., red cores) in low-density regions compared to high-density regions.

The environmental dependence of color gradients in quiescent red-sequence galaxies can be 
 understood as the following. 
Galaxy evolution proceeds differently depending on the environment. 
 When galaxies undergo merging events, the color gradients of progenitor galaxies (initially 
 red core) become diluted as a consequence of mix of stellar populations 
 (e.g., Ko \& Im 2005). 
Because the galaxies in high-density regions may have experienced more merging events, 
 the color gradients of galaxies there become gentler than those of galaxies in low-density 
 regions. 
If we assume that quiescent red-sequence galaxies with mid-IR and near-UV excess have 
 experienced centrally concentrated star formation within $\sim$2 Gyr, 
 much younger stellar populations would exist in the cores of galaxies.
This results in high values of color gradients (i.e., bluer core).  
However, the observed color gradients of quiescent red-sequence galaxies with near-UV and 
 mid-IR are excess are similar to those without near-UV and mid-IR excess.
This result indicates that the amount of young stars generated during recent star formation 
 is not large enough to change the color gradients, or that recent star formation is not 
 centrally concentrated. 
Because the near-UV and mid-IR colors are more sensitive to recent star formation than 
 optical colors, it would be interesting to study the optical$-$UV or optical$-$IR color 
 gradients in quiescent red-sequence galaxies. 
This will be discussed in a forthcoming paper. 
In contrast to quiescent red-sequence galaxies, E+A galaxies have positive color gradients 
(i.e., blue cores) and are more likely centrally concentrated (i.e., $c_{in}$ is smaller) than quiescent red-sequence galaxies, indicating recent, intense, centrally-concentrated starbursts (e.g., Yang et al. 2008).
 
%These different environmental trend of galaxies
%showing recent star formation implies a differnt galaxy evolution in various environment.

%% Fig. 8  ----------------------------------------------------------------
\begin{figure}
\centering
\includegraphics[width=12cm]{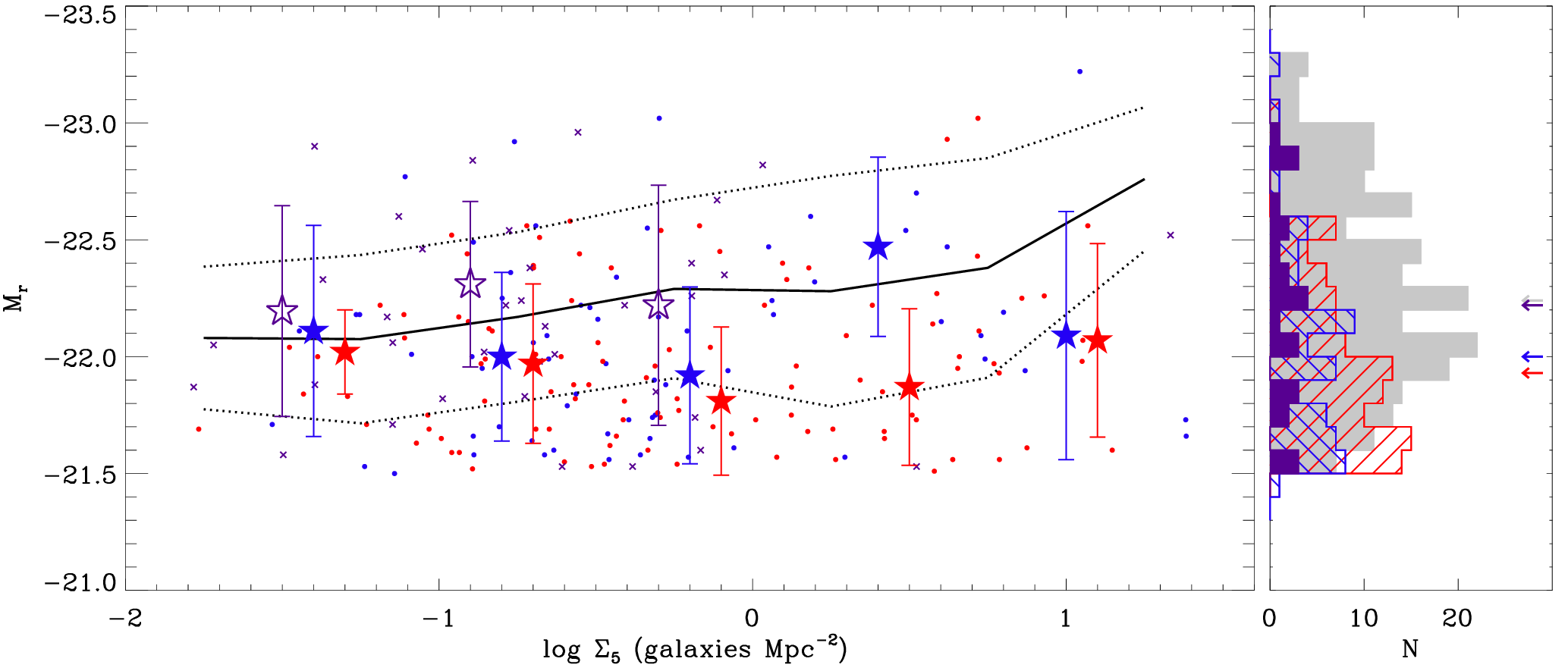}
\includegraphics[width=12cm]{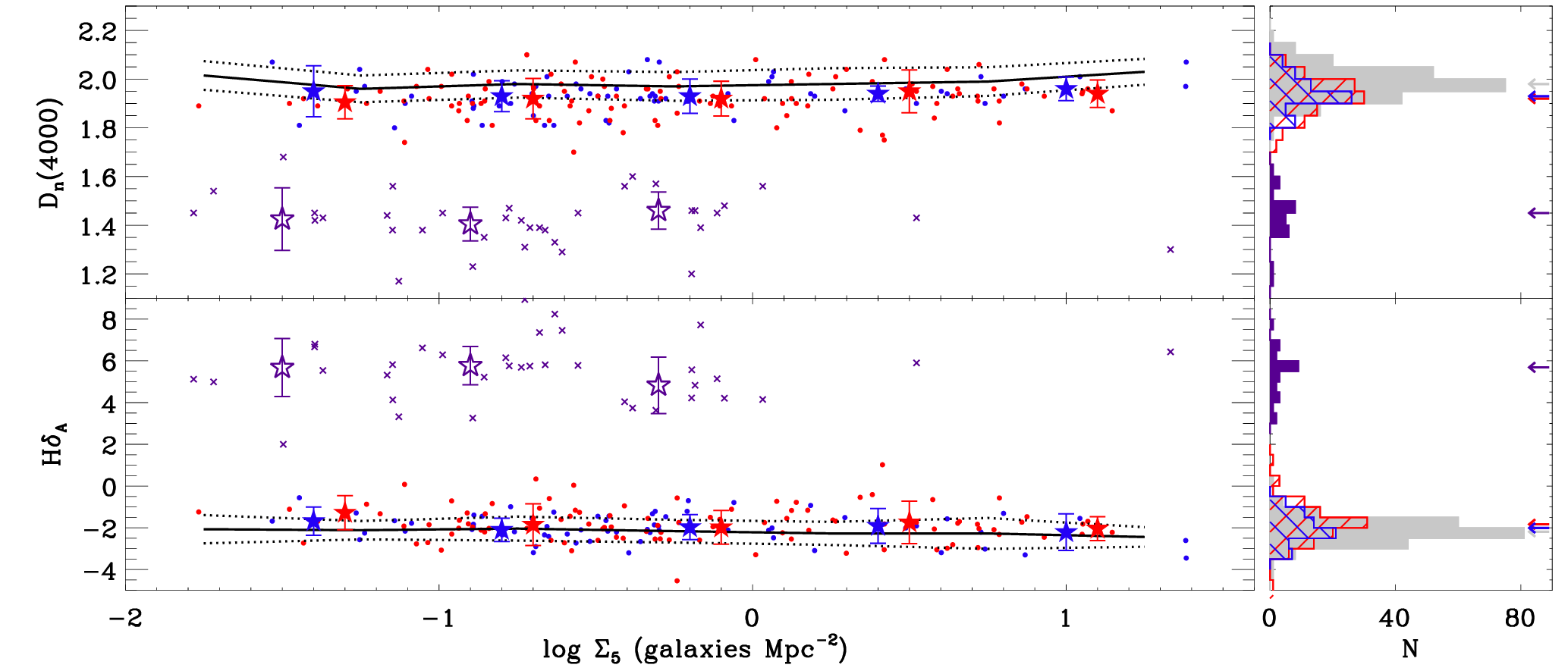}
\includegraphics[width=12cm]{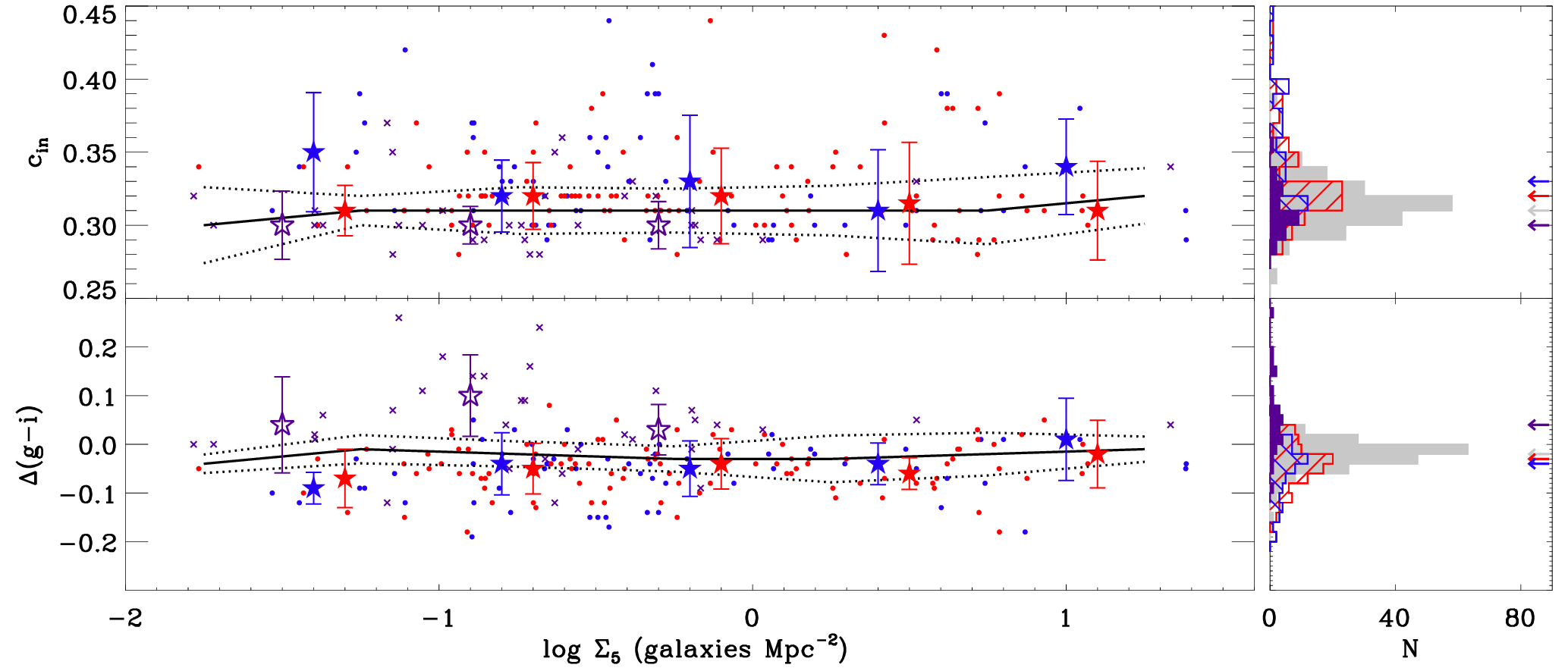}
\caption{Absolute magnitude $M_{r}$, $D_{n}$4000, H$\delta_{A}$, inverse concentration
         index $c_{in}$, and radial ($g-i$) color gradient as a function of local density 
         $\Sigma_{5}$.
         Red and Blue circles represent bright ($M_{r}$ $<$ $-$21.5), quiescent red-sequence
         galaxies with mid-IR and near-UV excess, respectively. 
         Red and blue star symbols are median values of those with mid-IR and near-UV excess 
         at each density bin, respectively.       
         Solid lines indicate median values of those without mid-IR and near-UV excess, 
         and two dotted lines represent 1 $\sigma$.
         For comparison, E+A galaxies are plotted with violet symbols 
         (crosses and open stars). 
         Density bins with less than three galaxies are excluded.
         Gray and violet filled histograms show quiescent red-sequence galaxies without
         mid-IR and near-UV excess and E+As, and quiescent red-sequence galaxies with mid-IR 
         and near-UV excess are denoted by hatched histograms with red and blue colors, 
         respectively.
         The arrows indicate the median values of each sample.}
\end{figure}

In summary, the comparison of quiescent red-sequence galaxies with mid-IR and those with 
 near-UV excess suggests that the two systems have similar physical properties. 
However, they are significantly different from E+A galaxies (i.e., post-starburst systems). 
Their physical properties are similar to those without mid-IR and near-UV excess, 
 but the objects with mid-IR and near-UV excess are younger than those without mid-IR and 
 near-UV excess.
Therefore, if we assume a simple recent star formation history (based on Figure 6), 
 near-UV excess galaxies with more than 10\% of young ($\lesssim$ 1 Gyr) component may 
 passively evolve into mid-IR excess galaxies within $\sim$1 Gyr (i.e., no longer near-UV 
 excess).
They then may gradually migrate into galaxies without mid-IR and near-UV excess later. 
The E+A galaxies may evolve into quiescent red-sequence galaxies with mid-IR excess within 
 $\sim$1 Gyr, if left alone.
On the other hand, near-UV excess galaxies with less than 10\% young component are likely to 
 directly evolve into quiescent red-sequence galaxies without mid-IR and near-UV excess 
 within $\sim$1 Gyr.
Moreover, this evolution among the quiescent red-sequence galaxies is expected to occur 
 earlier in massive galaxies.
% because NUV- and mid-IR-QRGs are slightly fainter (less massive) than p-QRGs.
 
The amount of AGB dust emission is larger for galaxies with younger ages because more massive 
 and luminous AGB stars are alive.
However, in principle, quiescent red-sequence galaxies without mid-IR and near-UV excess can 
 also have mid-IR excess emission from low-mass stars 
 even though the amount of mid-IR emission is small. 
This can suggest that the evolution from mid-IR excess galaxies to galaxies without mid-IR 
 excess probably takes a significant amount of time. 
A broad distribution of mid-IR colors (compared to near-UV colors) for quiescent red-sequence
 galaxies in Figure 6(a) seems to support this idea.

\subsection{Environmental Dependence of near-UV and mid-IR Excess Galaxies}

Figure 9(a) shows the distribution of local density ($\Sigma_{5}$) for the volume-limited 
 sample of quiescent red-sequence galaxies with $M_{r}$ $<$ $-$21.5 (dashed line).
We also plot all the galaxies with $M_{r}$ $<$ $-$21.5 and 0.04 $<$ $z$ $<$ 0.11 
 (gray histogram).
The histogram for quiescent red-sequence galaxies shows an excess in high-density regions 
 compared to all the galaxies in the volume. 
In the middle and bottom panels, we plot the fractions of near-UV and mid-IR excess galaxies  
 among the quiescent red-sequence galaxies as a function of local density. 
We divide the galaxy sample into two groups based on their $r$-band absolute magnitudes so 
 that each subsample has the same number of galaxies.

The middle panel (b) shows that the fraction of near-UV excess galaxies is large in low-  
 density regions. 
This environmental dependence seems stronger for bright galaxies (filled circles). 
The figure also shows that there are a significant number of bright galaxies with near-UV 
 excess (i.e., recent star formation at $\lesssim$ 1 Gyr) in the lowest density bin. 
On the other hand, faint galaxies (open circles) show a high fraction of near-UV excess 
 galaxies in the highest density bin. 
It should be noted that density bins with less than three galaxies are excluded. 
Actually, the fraction of bright galaxies in the highest density bin is 8\% (2 out of 26). 
This environmental dependence of near-UV excess galaxies is consistent with the results of 
 Schawinski et al. (2007). 

%While the luminosity dependence in the high-density environment is quite evident, 
%we only count neighbors with $M_{r}$ $\leq$ $-$20.66, so large number of faint
%neighbors could be missed.
%(which are likely to be merged into our bright ($M_{r}$ $\leq$ $-$21.5) galaxies)
 
On the other hand, the fraction of mid-IR excess galaxies appears to increase with local  
 density (see panel (c)). 
The enhancement of mid-IR excess galaxies in high-density regions is mainly because of faint 
 galaxies (open circles); roughly a half of faint, quiescent red-sequence galaxies are mid-IR 
 excess galaxies (i.e., recent star formation at $\lesssim$ 2 Gyr). 
The luminosity dependence of the fraction of mid-IR excess galaxies is consistent with the 
 results in Ko et al. (2009, 2012); faint (low mass) galaxies are more likely to have mid-IR 
 excess emission among red early-type galaxies. 
However, the environmental dependence of the fraction of mid-IR excess galaxies (i.e., higher fraction of mid-IR excess galaxies in high-density regions than in low-density regions) seems different from our previous results 
(i.e., low fraction of red early-type galaxies with mid-IR excess emission in high-density regions). 
However, this difference results from a different sample selection between the two studies. 
%In other words, the mid-IR excess galaxies in Ko et al. (2009, 2012) are divided into two 
%subsamples of quiescent red-sequence galaxies with near-UV and mid-IR excess in this work.
Ko et al. (2009, 2012) used only the mid-IR excess galaxies, but in this study we
 further divide the mid-IR excess galaxies into two subsamples based on near-UV excess.
If we consider mid-IR excess galaxies without near-UV excess cut, the fraction of mid-IR excess galaxies are $\sim$58\% and $\sim$25\% in the lowest and highest density bins, respectively. 
We then have a lower fraction of quiescent red-sequence galaxies with mid-IR excess in high-density regions than in 
low-density regions, consistent with our previous results.

In summary, there is a strong environmental dependence of fractions of near-UV and mid-IR 
 excess galaxies, indicating that the recent star formation activity of quiescent 
 red-sequence galaxies are strongly affected by the environment. 
These results can suggest a possible scenario of recent (at least $\sim$2 Gyr) evolutionary 
 history of bright early-type galaxies, migrating to the red sequence with quiescent mode. 
We can speculate that recent (within $\sim$1 Gyr) star formation of quiescent red-sequence galaxies (traced by near-UV excess) occurs 
 in any environments, but preferentially in low-density regions at current epoch. 
Recent star formation in quiescent red-sequence galaxies is also common irrespective of their luminosities 
 except bright near-UV excess galaxies in the highest density region (e.g., galaxy cluster
 regions). 
The bright, quiescent red-sequence galaxies in cluster regions seem to have no gas left to induce recent star formation or to have less frequent 
 galaxy mergers, so they are not detected in the near-UV. 
However, the brightest cluster galaxies in the center of cluster can have recent star formation because of
 cooling intracluster gas (e.g., Hicks et al. 2010). 

Quiescent red-sequence galaxies that might have experienced star formation $\gtrsim$ 1 Gyr 
 ago (traced by mid-IR excess) can also exist in any environments. 
However, they are found preferentially in high-density regions and most bright mid-IR excess 
 galaxies are rare in low-density regions.
The high-density regions including galaxy clusters keep accreting galaxies, and the crossing time 
 for a galaxy in clusters is about a few Gyrs (Treu et al. 2003; Park \& Hwang 2009). 
The near-UV excess galaxies may evolve into mid-IR excess galaxies within $\sim$1 Gyr and remain as mid-IR excess galaxies for 0.5$-$1 Gyr 
 if left alone. 
These can suggest that the mid-IR excess galaxies in high-density regions are descendants of near-UV excess galaxies recently 
 accreted from low/intermediate density regions.

%Galaxies at high density have been undergoing very recent SF a much higher frequency of minor mergers 
%(triggering post-starburst) at the low-density region. Subsequently, their descendants are infalling
%into the intermediat- and high-density regions. 
  
%This implies that environmental dependence of mid-IR-excess galaxies among QRGs varies with mass. 
%In other words, recent star formation history (within $\sim$2 Gyr) is strongly dependent on the mass, 
%and also secondly affected by the environment.

%We find that the NUV- and mid-IR-QRGs fraction is both dependent on luminosity in that fainter 
%galaxies have higher fraction of NUV- and mid-IR-excess galaxies. This is largely consistent with 
%previous studies (e.g., Ko et al. 2012), suggesting that a mass-dependent star formation history 
%where massive galaxies are much older and became passive earlier than less massive galaxies. 

%% Fig. 9  ----------------------------------------------------------------
\begin{figure}
\centering
\includegraphics[width=15cm]{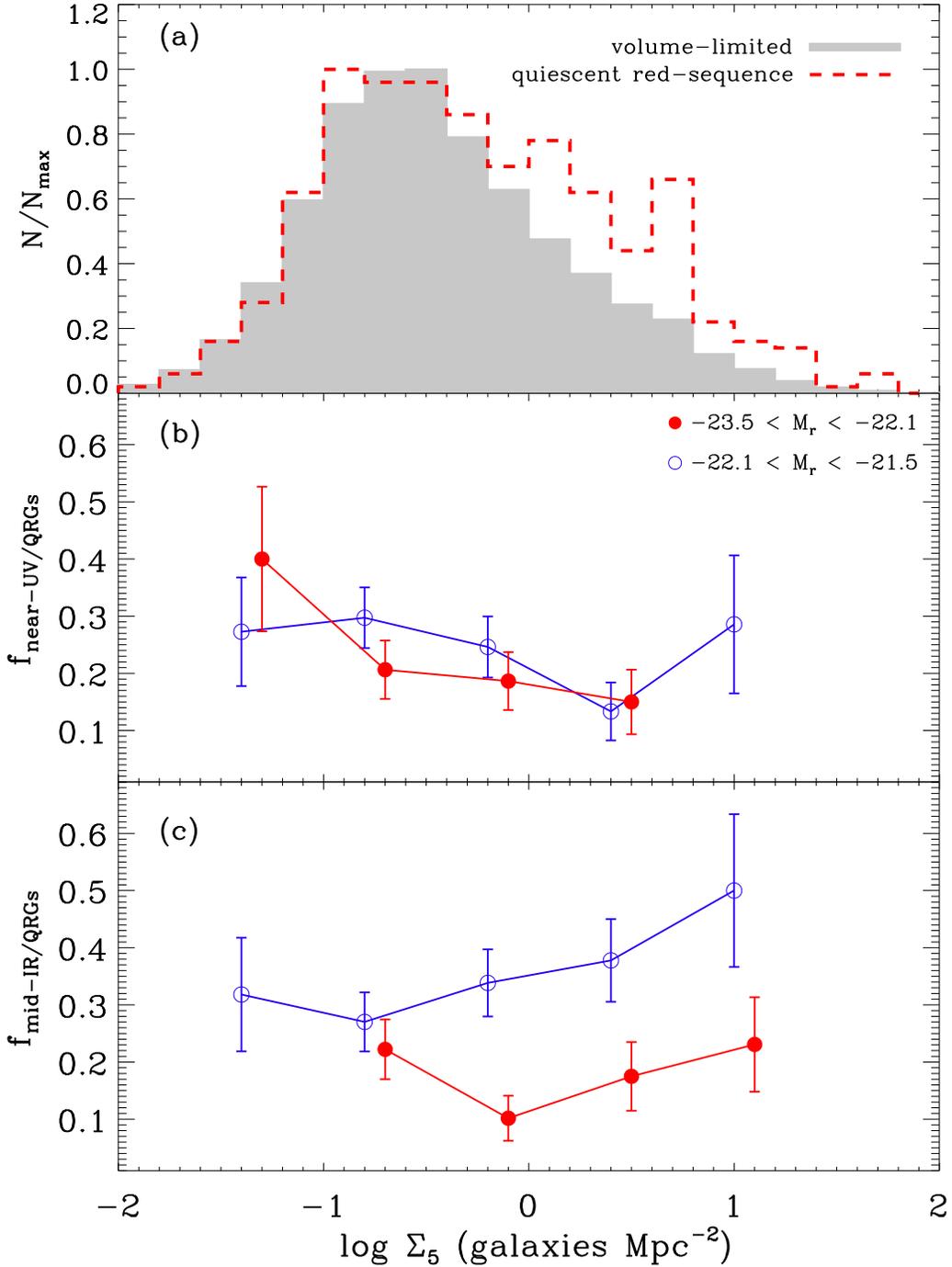}
\caption{Distribution of local density $\Sigma_{5}$ for bright ($M_{r}$ $<$ $-$21.5), 
         quiescent red-sequence galaxies 
		 (dashed line) and all the galaxies with $M_{r}$ $<$ $-$21.5 and 
		 0.04 $<$ $z$ $<$ 0.11 (filled histogram) (a). 
         Fractions of near-UV excess galaxies (b) and of mid-IR excess galaxies (c) 
         among bright ($M_{r}$ $<$ $-$21.5), quiescent red-sequence galaxies.  
         Density bins with less than three galaxies are not plotted.}
\end{figure}
%% ------------------------------------------------------------------ end Fig. 9

\section{SUMMARY AND CONCLUSIONS}

We use {\it WISE} mid-IR and {\it GALEX} near-UV data for a spectroscopic sample of SDSS 
 galaxies at 0.04 $<$ $z$ $<$ 0.11 and $m_{r}$ $<$ 17.77 to study the recent star formation 
 history of bright ($M_{r}$ $<$ $-$21.5), early-type galaxies on the optical red sequence. 
Among the 648 quiescent red-sequence galaxies, 55\% show mid-IR excess emission over the 
 stellar component. 
Because we select them not to have AGNs, H$\alpha$ emission, and highly inclined disks, 
 the mid-IR emission mainly results from the circumstellar dust around AGB stars or 
 from the stellar photosphere.

%The QRGs with NUV and mid-IR excess show signs of RSF with little ongoing SF, but their
%optical red colors mainly result from their underlying old stellar populations.
%This suggests that they have just arrived to the red sequence after RSF activity.

Among the 648 quiescent red-sequence galaxies, 26\% show near-UV excess emission,
 indicating recent star formation within $\sim$1 Gyr. 
If we consider bright ($M_{r}$ $<$ $-$21.5) early-type galaxies, the fraction is down to 14\%. 
%Compared to the result of Schawinski et al. (2007), our selection of QRGs showing RSF
%is very conservative. 
%our bright QRGs are corresponding to the sample of previous NUV studies of ETGs 
%a half of the previous result of Schawinski et al. (2007).
%The fraction of quiescent red-sequence galaxies with mid-IR excess emission is 55\%. 
Most near-UV excess galaxies also show mid-IR excess emission, 
 but some mid-IR excess galaxies do not show near-UV excess.
This is consistent with that the mid-IR light is sensitive to star formation
 over longer time scales (i.e., recent star formation within $\sim$2 Gyr) than the near-UV light. 
If we consider bright early-type galaxies, the fraction of quiescent red-sequence galaxies 
 with recent star formation is 39\%.
This suggests that the recent star formation is common among nearby, quiescent, red, 
 early-type galaxies.

The near-UV and mid-IR excess galaxies (i.e., early-type galaxies with recent star formation) are optically fainter (or less massive) than galaxies without near-UV and mid-IR excess (i.e., early-type galaxies without near-UV and mid-IR excess emission). 
The mid-IR weighted mean stellar age and $D_{n}$4000 indicate that the near-UV and mid-IR excess galaxies are also 
 slightly younger than galaxies without near-UV and mid-IR excess. 
This can suggest an evolutionary scenario of quiescent red-sequence galaxies: 
 those with near-UV excess $\rightarrow$ those with mid-IR excess $\rightarrow$ those without
 near-UV and mid-IR excess.
%Although this evolution scenario depends on the amount of RSF,
% this could potentially give us new constraints for understanding recent star formation history 
% of bright, early-type, red-sequence galaxies.
The near-UV and mid-IR excess galaxies show different environmental dependence; 
 near-UV excess galaxies are preferentially found in low-density regions, 
 but mid-IR excess galaxies are found in high density regions. 
This suggests strong environmental effects on the evolution of quiescent red-sequence galaxies 
 with recent star formation.
 
This study provides useful observational constraints on the theoretical models about the AGB stars 
 and their contribution to the mid-IR fluxes in galaxies. 
However, we do not quantify the amount of mid-IR emission attributed to AGB dust in galaxies, 
 and do assume that the contribution from other mechanisms for mid-IR emission 
 (e.g., low-level of dusty starburst, AGNs) is small.   
These will be examined in a forthcoming paper. 
%It should be noted that we need to quantify how much fraction of mid-IR emission of a galaxy 
 % is attributed to AGB dust, even though the other mechanisms (e.g., low-level of SF and AGNs)
 %are considered to be removed in this work. 

\begin{acknowledgements}

We thank the anonymous referee for useful comments that greatly improved this paper.
J.K. and J.C.L. are the members of Dedicated Researchers for Extragalactic AstronoMy (DREAM) 
in Korea Astronomy and Space Science Institute (KASI).
HSH acknowledges the Smithsonian Institution for the support of his post-doctoral fellowship.
We warmly thank Yumi Choi, Hyun-Jin Bae and Suk-Jin Yoon for providing their E+A galaxy catalog.

This publication makes use of data products from the Wide-field Infrared Survey Explorer, which is a joint project of the University of California, Los Angeles, and the Jet Propulsion Laboratory/California Institute of Technology, funded by the National Aeronautics and Space Administration.
{\it GALEX} is a NASA Small Explorer, launched in 2003 April.
We gratefully acknowledge NASA's support for construction,
operation, and science analysis for the 
{\it GALEX} mission, developed in cooperation with the Centre National d'Etudes Spatiales of France and the Korean Ministry of Science and Technology.
Funding for the SDSS and SDSS-II has been provided by the Alfred P. Sloan Foundation, the Participating Institutions, the National Science Foundation, the US Department of Energy, the National Aeronautics and Space Administration, the Japanese Monbukagakusho, the Max Planck Society, and the Higher Education Funding Council for England. The SDSS Web Site is http://www.sdss.org/. The SDSS is managed by the Astrophysical Research Consortium for the Participating Institutions. The Participating Institutions are the American Museum of Natural History, Astrophysical Institute Potsdam, University of Basel, Cambridge University, Case Western Reserve University, University of Chicago, Drexel University, Fermilab, the Institute for Advanced Study, the Japan Participation Group, Johns Hopkins University, the Joint Institute for Nuclear Astrophysics, the Kavli Institute for Particle Astrophysics and Cosmology, the Korean Scientist Group, the Chinese Academy of Sciences (LAMOST), Los Alamos National Laboratory, the Max-Planck-Institute for Astronomy (MPIA), the Max-Planck-Institute for Astrophysics (MPA), New Mexico State University, Ohio State University, University of Pittsburgh, University of Portsmouth, Princeton University, the United States Naval Observatory, and the University of Washington.

\end{acknowledgements}

\clearpage

\end{document}